# Cognitive Radars: A Reality?

Maria S. Greco[1], Fulvio Gini[1], Pietro Stinco[1], Kristine Bell[2]

[1]Dipartimento di Ingegneria dell'Informazione, University of Pisa, Italy
[2]Metron, Inc., USA
E-mails: { f.gini,m.greco}@iet.unipi.it

**Abstract—** This paper describes some key ideas and applications of cognitive radars, highlighting the limits and the path forward. Cognitive radars are systems based on the perception-action cycle of cognition that sense the environment, learn from it relevant information about the target and the background, then adapt the radar sensor to optimally satisfy the needs of their mission according to a desired goal. The concept of cognitive radar was introduced originally for active radar only. In this paper we describe how this paradigm can be applied also to passive radar. In particular, we describe (i) cognitive active radars that work in a spectrally dense environment and change the transmitted waveform on-the-fly to avoid interference with the primary user of the channel, such as broadcast or communication systems, (ii) cognitive active radars that adjust transmit waveform parameters to achieve a specified level of target tracking performance, and (iii) cognitive passive radars, that contrary to the active radars cannot directly change the transmitted waveforms, but can instead select the best source of opportunity to improve detection and tracking performance.

## 1. Introduction

The idea of cognitive radar was introduced for the first time by S. Haykin in 2006 [Hay06]. However, the first papers on knowledge-based systems and agile waveform design, which are the foundations of the modern concept of cognitive radar, trace back to the late 90's (see [Gin08], [Aub13], [Blu16] and references therein). Quoting [Hay06], a cognitive radar *"continuously learns about the environment through experience gained from interactions with the environment, the transmitter adjusts its illumination of the environment in an intelligent manner, the whole radar system constitutes a dynamic closed feedback loop encompassing the transmitter, environment, and receiver"*.

The new feature of a cognitive radar that differentiates it from a classical radar is the active feedback between receiver and transmitter, as shown in the block diagram of Fig. 1; the classical concept of adaptivity, already known in the radar community since the early 60's, is extended to the transmitter. A classical adaptive radar is able to extract information from the target and the disturbance signals through appropriate signal processing algorithms and to use that information at the receive level to improve its performance. A cognitive radar conversely is able to use all the extracted information not only at the receive level but also at the transmit level by changing on-the-fly the transmit frequency channel, waveform shape, time-on-target, pulse repetition frequency (PRF), power, number of pulses, polarization, etc. In an adaptive radars all these parameters are pre-set and cannot be changed on-the-fly.

The cognitive radar system mimics the perception-action cycle of cognition [Fus03], [Hay12]. It senses the environment and learns from it important information about the target and the background (perception), then adapts the transmitted waveform to optimally suit the needs of its mission (surveillance, tracking, etc.) according to a desired goal (action). In this "decision-action" phase, there are two main approaches that can be applied: (i) the Bayesian approach, which builds on prior distributions and knowledge-aided models of the environment obtained from past measurements in the same or similar environments [Hay 12], and (ii) the machine learning approach, which determines the next action based only on the measured data and knowledge of actions commonly taken in the same or similar environments [Met15].

Figure 1 - Block diagram of cognitive radar seen as a dynamic closed-loop feedback system with the perception-action cycle [Hay12].

In nature, all desired features of a cognitive radar system are embedded in the echo-location systems of bats and dolphins. It is well known (see e.g. [Sim73], [Tho04]) that both bats and dolphins are able to "see" very small prey (compared to their own size) and can track them by adjusting both the duration and the repetition frequency of their emitted pulse bursts based upon the range and the velocity of the targets. Some dolphin species, as the Bottlenoses, are able to detect, classify, and localize targets the size of a sardine, in a cluttered background, over ranges from 0 m to about 150 m, in any sea condition, and any maritime environment, from the open ocean to rivers and estuaries [Gre06]. Of course, it is not trivial to implement in a real radar or sonar system all the functionality of a well-trained bio-sonar, which has been evolving over millions of years, although some efforts along these lines have been successfully pursued [Ves09].

In this paper, we describe some applications of cognition to both active and passive radars, highlighting the limitations and the path forward. The remainder of this paper is organized as follows. In Sections II and III, the need for a new cognition paradigm in active radar systems is motivated by the increasing erosion of the spectrum portion dedicated to them and by the ongoing need to improve detection and tracking performance. In Section IV, we describe where cognition can be applied in passive systems, where the sensing is adapted by choosing the best transmitter of opportunity. Conclusions are drawn in Section V.

## 2. Cognitive active radars that adapt to the environment: the evil of the spectrum erosion

Radar technology has recently been evolving towards higher resolution, high-precision multifunction systems with an ever-increasing list of capabilities, all available simultaneously, such as surveillance, tracking, confirmation of false alarm, back-scanning, and clutter and interference estimation. These capabilities are traditionally associated with dedicated individual radars [Wic10]. For these reasons, multifunction radar systems should be able to work with frequency bands wider than traditional ones. Clearly, this is in conflict with the growth of activities in the area of civil communications, where the emergence of new technologies and new services that have a high demand for spectrum allocation puts a very strong pressure upon the frequency channels currently allocated to radars. The allocation of spectrum is regulated by the International Telecommunication Union (ITU) and is continually reviewed at an international level by the World Radiocommunication Conference (WRC) [Gri15]. Some portions of the radar bands have been recently allocated to communication services. For instance, in the US, the National Telecommunications and Information Administration (NTIA) [NTI13], [ECC16] has recently devoted efforts to identifying frequency bands that could be made available for wireless broadband service provisioning. A total of 115 MHz of additional spectrum (1695-1710 MHz and 3550-3650 MHz bands) has been identified for wireless broadband systems [USD10]. Moreover, high UHF radar systems overlap with GSM communication systems and S-band radars already partially overlap with Long Term Evolution (LTE) and WiMax systems [Dar13]. Some results on the impact of S-band radars on a WiMax systems are shown in [Coh10] and the impact of VHF/UHF radars on DVB-T and DVB-T2 systems has been studied in [Mel12] as a function of modulation scheme, propagation environment, and radar waveform type.

From the examples above, it is evident that the availability of frequency spectrum for multifunction radar systems has been severely compromised and the available frequency bands are increasingly shrinking. In the near future, radar systems will have to share their bandwidth with communications systems [Rom15], [Den13], [Li16] [Zhe18], where the latter are quite often the primary users. It is very likely that they will also share the same platforms and the same antennas in dual-function radar-communications systems [Has16]. It is clear that this issue of spectrum crowding cannot be addressed only by traditional modes of operation, such as spatial signal processing and beamforming [Far92]. Future systems require the ability to anticipate the behavior of emitters in the operational environment and to adapt their transmissions in a cognitive fashion based upon the spectrum availability.

To further improve the efficiency of spectrum utilization by the radars, modern systems should also be able to change the transmitted waveform on-the-fly (adaptive radar illumination or waveform diversity). Again, the radar should apply its cognition to extract, from the past observed radar returns, useful information in order to select or decide the waveform for next transmission [Gin12], [Blu16], [Bel15], possibly predicting the consequences of its actions, and using memory to store the learned knowledge [Hay06].

In this section, we describe some examples of cognitive methodology used to reduce mutual interference between the radar and other radiating elements. Radar cognition in this scenario has two main components: perception in the form of spectrum sensing and action in the form of spectrum sharing. The goal of spectrum sensing is to recognize the frequencies used by other systems using the same spectrum in real time, while the goal of spectrum sharing is to design the transmitted waveform to limit interference between the radar and other services.

In this case study, the transmitting systems other than the radar are considered the primary users, that is, the users who have higher priority or legacy rights on the usage of a specific part of the spectrum. The radar is a secondary user, which has lower priority and must exploit this spectrum in such a way that it does not cause interference to the primary users. There is not any cooperation between the radar and the communication systems. Therefore, the radar needs cognitive radio capabilities [Den13], such as sensing the spectrum reliably to check whether a primary user is using it and to change the transmission parameters to exploit the unused part of the spectrum. As an illustrative example, Fig. 2 shows the spectrum opportunities in the frequency channels [Gre16].

**Figure 2** – Spectrum opportunities [Gre16].

The available spectrum is divided into narrower chunks of bands. Spectrum opportunity means that all the bands are not used simultaneously at the same time, therefore some bands might be available for opportunistic usage. The block diagram of the analyzed cognitive system [Sti16a], [Sti16b] is plotted in Figure 3.

**Figure 3** – Blocks in a cognitive radar.

The first two blocks, "Compressive Sensing (CS) Spectrum Reconstruction" and "Energy Detector (ED)" perform the spectrum sensing function of the radar system. All the frequency channels are scanned and the compressive sensing technology is used to reconstruct the instantaneous amplitude of the signal in each channel while the energy detector is used to decide the occupancy of a channel. Under the hypothesis that the frequency spectrum is sparse (only few channels are occupied at the same time), CS can be profitably used to solve the problem of hardware constraints by reducing the sampling rate and decreasing the computational complexity [Gao16], [Qin15].

In this particular example, the Pulse Repetition Interval (PRI) of the radar system and the time slot duration of the communications system are assumed to be of the same length. This is not always true in practical scenarios. We considered it as the worst case. Therefore, for each frequency channel at the time of transmission (i.e. at the beginning of each PRI), the radar is not able to measure whether the frequency channel is effectively occupied by the communication system (the available time is too short). The information coming from the ED, concerning the behavior of the primary users, is instead continuously saved and analyzed in order to estimate the parameters characterizing the primary user models and to evaluate the probability of having a spectrum opportunity, that is, the probability that the monitored frequency channel is free at the time of transmission. If this probability is sufficiently high, the cognitive radar transmits, otherwise if the probability is too low, the radar does not transmit.

Of course, the ED is not the only method which can be applied to monitor the channel occupancy. For cognitive radios, many other methods have been proposed such as waveform or matched filter-based detectors, feature-based detectors, and some emerging eigenvalue-based and wavelet-based detectors, as well documented in [Yuc09]. Each of these detectors has its advantages and disadvantages with varying detection capabilities, implementations and complexities, sensing times, and assumptions and requirements on the primary user signal. Recent advances on spectrum exploitation and exploration technique can be found in [Lun15], as well as in [Mas09], [Mel15], [Mod17].

For predicting the occupancy of the channels by primary users, Hidden Markov Models (HMMs) have been proposed in the literature (see e.g. [Bar11] and [Rab89]). In [Sti16a], a very easy HMM for the primary users was applied, as shown in Fig. 4.

**Figure 4** - Hidden Markov Model representation for spectrum occupancy.

In general, an HMM is comprised of a set $S_t$ of possible states and a set $O_t$ of possible emissions. The possible states represent the real activity of the primary user in each frequency channel. If the primary user is transmitting at time slot $t$, the state is $S_t=1$, otherwise, if the channel is free, the state is $S_t=0$. However, due to the noise in the channel, a free channel can be mis-classified as busy and a busy channel mis-classified as free. Therefore, there are also two possible emissions, which are represented by the observation symbol $O_t$ at the output of the spectrum sensing detector.

The primary user's dynamic, described by the states $S_t=0$ and $S_t=1$, is characterized by the 2×2 state transition probability matrix **A**, which represents the probabilities associated with changing from one state to another and is given by $[\mathbf{A}]_{hk} = a_{hk} = \Pr[S_t = h | S_{t-1} = k]$, with $h,k=\{0, 1\}$. The transitions from the states $S_t$ to the observations $O_t$ are described by the 2×2 emission probability matrix **B**, which represents the probabilities associated with obtaining a certain output given that the model is currently in a true state $s_t$. Hence, $[\mathbf{B}]_{hk} = b_h(k) = \Pr[O_n = h | S_n = k]$.

The emission probability matrix **B** is related to the Receiver Operating Characteristic (ROC) of the Spectrum Sensing detector, the ED in the scheme at hand. As a matter of fact, $b_0(1)$ is the probability of false alarm, that is the probability of classifying a free channel as busy, whereas $b_1(0)$ is the probability of missed detection, that is the probability of classifying a busy channel as free. The other parameters characterizing the primary users are the initial state distribution $\pi=\{\pi_i\}$, defined as $\pi_i = \Pr[s_1 = S_i]$, with i=0,1. The matrix **B** is generally assumed to be known (or estimated through Monte Carlo simulation). Hence, the problem of Channel Parameter Estimation, in the third block of Fig. 4, is to determine a method to estimate the model parameters **A** and **π** using a finite observation sequence **O**=[$O_1$...$O_T$] of $T$ elements, provided by the ED. It is impossible to solve this problem in closed form [Rab89], however the solution can be found with iterative procedures. The most widely adopted procedure is the Baum-Welch (BM) method [Bau68], which is closely related to Expectation-Maximization (EM) [Rab89], [Dem77]. The details of the Baum-Welch method, as applied to cognitive radar, are described in [Sti16a]. The matrix **A**, the probability $\gamma(0)$ that in the previous PRI the channel was free, and the probability $\gamma(1)$ that it was busy, are updated continuously at each PRI by the BM algorithm using the new observations coming from the ED (the history of the channel).

The probability $p_{SO}$ of having a spectrum opportunity is then given by:

$$p_{SO} = \gamma(0)a_{00} + \gamma(1)a_{01} \quad (1)$$

where $\gamma(0)$ and $\gamma(1)$ have been already defined and $a_{00}$ and $a_{01}$ are the probability that in the current PRI the channel remains free and the probability that in the current PRI it becomes free, respectively. It is calculated in the fourth block in Fig. 4.

The last block of the scheme in Fig. 4 compares the $p_{SO}$ with a threshold λ, and transmits only if the probability is greater than the threshold. Clearly, there are two kinds of errors. The first error event $e_0$ is when the cognitive radar does not transmit and the channel is free, that is the probability of losing a spectrum opportunity. The other error event $e_1$ is when the radar transmits and the channel is occupied by the primary user, that is the probability of having a collision. Fig. 5 shows the probability of these two errors as a function of the threshold λ [Sti16a]. This graph can be used to tune the cognitive radar to the desired performance. It is clear that when λ=0, the radar is always transmitting, therefore the probability of $e_1$ is the same as the probability that the channel is busy, which was set to 0.5 for this example.

Similarly, when λ=1, the radar never transmits and the probability of $e_0$ coincides with the probability that the channel is free. Fig. 6 shows the probability of losing a spectrum opportunity and the probability of having a collision as a function of time, obtained by observing the performance of the system for about 10,000 PRI's. These results have been obtained through $10^3$ Monte Carlo runs, with λ =0.65 [Sti16a].

**Figure 5** – Probabilities of $e_0$ and $e_1$ as a function of λ [Sti16a].

**Figure 6** - Probabilities of $e_0$ and $e_1$ as a function of time [Sti16a].

The simulation results show that the error probabilities of the cognitive radar that adopts the proposed methodology are constant over time and much lower than that of a classical radar that always transmits, ignoring the presence of the primary user, and lower than that of a radar that never transmits to avoid interference with the primary user.

To further increase the spectrum awareness of a cognitive system (radio or radar), it has been proposed to use a Radio Environmental Map (REM) [Zha07]. The idea behind the REM is to store and process a variety of data to extract all the available information on transmitter locations, propagation conditions, and spectrum usage in space and time.

Exploiting the REM, the radar could become aware of the surrounding electromagnetic environment, and then intelligently use the transmit bandwidth and probing waveforms [Aub16].

## 3. Cognitive active radars that adapt to the target

Radar sensors are the first stage in sensor/processor systems involved in detection, localization, tracking, and classification. These functions can be improved via adaptation of the sensor waveform and radar system parameters using feedback from the output of the end processor. In this section, we describe how Bayesian filtering of the target state can be used to implement the perception portion of the perception-action cycle, and discuss information measures that can be used to optimize the next transmission (action) of the radar. The full mathematical details can be found in [Bel15]. This work generalizes and formalizes the work of Haykin and others in [Hay06], [Hay10], [Hay12], [Ker94], [Kre05a], [Kre05b], [Kre07], [Hur08], [Sir09], [Cha12], [Cha15], and the references given in [Bel15].

The mathematical model that describes the cognitive sensor/processor system is illustrated in Fig. 7. The system consists of five components: (i) the Scene, which includes the target and environment, (ii) the Sensor, which observes the scene either through active probing or passive observation, (iii) the Processor, which converts the observed data into a perception of the scene, and (iv) the Controller, which determines the next actions taken by the sensor and/or processor.

**Figure 7** – Cognitive Sensor/Processor System

We assume that the objective is to estimate the state of a target at time $k$, which is denoted as $\mathbf{x}_k$. The sensor observes the scene and produces a measurement vector $\mathbf{z}_k$ which depends on the target state $\mathbf{x}_k$ and the sensor parameters $\boldsymbol{\theta}_k$. The estimate of the target state at time $k$ will be a function of the observations up to time $k$, which in turn depend on the sensor parameters up to time $k$, which are denoted as $\mathbf{Z}_k = \{\mathbf{z}_1, \mathbf{z}_2, ..., \mathbf{z}_k\}$ and $\boldsymbol{\Theta}_k = \{\boldsymbol{\theta}_1, \boldsymbol{\theta}_2, ..., \boldsymbol{\theta}_k\}$, respectively.

Our perception of the target state is modelled probabilistically by a target state probability density function (pdf). The initial target state pdf is denoted as $q(\mathbf{x}_0)$ and the transition pdf, which represents the probability that a target in state $\mathbf{x}_{k-1}$ will evolve to state $\mathbf{x}_k$, is denoted as $q(\mathbf{x}_k \mid \mathbf{x}_{k-1})$. The measurements are modelled probabilistically by the conditional pdf $f(\mathbf{z}_k \mid \mathbf{x}_k; \boldsymbol{\theta}_k)$, which is also called the likelihood function. It depends on the target state and the sensor parameters used to obtain the measurement vector. The cost of obtaining an observation and any constraints on the sensor parameters are modeled by the sensor cost function $R_\Theta(\boldsymbol{\theta}_k)$. The processor maintains the current perception of the target state in the posterior pdf, which is the conditional pdf of the target state given all of the measurements, and is denoted as $f^+(\mathbf{x}_k) \equiv f(\mathbf{x}_k \mid \mathbf{Z}_k; \boldsymbol{\Theta}_k)$. The posterior pdf is computed from the Bayes-Markov recursions of Bayesian filtering [Sto14], [Ris04a]. These are commonly implemented using a variant of the Kalman Filter or a particle filter. The processor also produces an estimate of the target state $\hat{\mathbf{x}}_k(\mathbf{Z}_k)$ by minimizing the expected value of the processor cost function $C(\hat{\mathbf{x}}_k(\mathbf{Z}_k), \mathbf{x}_k)$.

The controller decides on the next value for the sensor parameters $\boldsymbol{\theta}_k$ by minimizing a loss function $L_{C,\Theta}(\cdot)$ that balances the processor performance via the processor cost function $C(\cdot,\cdot)$ and the cost of using the sensor via the sensor cost function $R_\Theta(\cdot)$. In the controller, we assume that we have received the observations up to time $k$-1 and want to find the next set of sensor parameters $\boldsymbol{\theta}_k$ to optimize the performance of the state estimator that will include the next observation $\mathbf{z}_k$ as well as the previous observations $\mathbf{Z}_{k-1}$. In [Bel15], we define the joint conditional pdf of $\mathbf{x}_k$ and $\mathbf{z}_k$ conditioned on $\mathbf{Z}_{k-1}$ as $f^\uparrow(\mathbf{x}_k, \mathbf{z}_k; \boldsymbol{\theta}_k) \equiv f(\mathbf{x}_k, \mathbf{z}_k \mid \mathbf{Z}_{k-1}; \boldsymbol{\Theta}_k)$, and define the predicted conditional Bayes risk as the expected value of the processor cost function with respect to the joint conditional pdf $R_C^\uparrow(\boldsymbol{\theta}_k) \equiv E_k^\uparrow\{C(\hat{\mathbf{x}}(\mathbf{Z}_k), \mathbf{x}_k)\}$. The next value of $\boldsymbol{\theta}_k$ is chosen to minimize a loss function that balances the predicted conditional Bayes risk and the sensor cost, $L_{C,\Theta}(\boldsymbol{\theta}_k) = L\{R_C^\uparrow(\boldsymbol{\theta}_k), R_\Theta(\boldsymbol{\theta}_k)\}$. The controller optimization problem is then given by:

$$\boldsymbol{\theta}_k^{opt} = \arg\min_{\boldsymbol{\theta}_k} L_{C,\Theta}(\boldsymbol{\theta}_k). \quad (2)$$

The cognitive sensor/processor system framework described by Fig. 7 is very general and can be applied to many problems. The framework can be specialized to single target tracking systems by defining the processor cost function to be the squared estimation error. The optimal target state estimator is then the mean of the posterior pdf [Van13], and the predicted conditional Bayes risk is the trace of the predicted conditional mean square error (PC-MSE) matrix. In most cases, it is not possible to evaluate the PC-MSE matrix analytically or numerically. However, we can use the Bayesian Cramér-Rao lower bound (BCRLB), which provides a (matrix) lower bound on the MSE matrix of any estimator [Van13], [Van07] and is usually analytically tractable. It is frequently used as a tool for system analysis in place of the MSE matrix. For tracking applications, application of the BCRLB theory yields the posterior Cramér-Rao lower bound (PCRLB) [Tic98]. The PCRLB provides a lower bound on the global MSE that has been averaged over $x_k$ and $Z_k$, thus it characterizes tracker performance for all possible data that might have been received. In [Bel15], we developed a predicted conditional Cramér-Rao lower bound (PC-CRLB) to bound the PC-MSE matrix, which is averaged over the joint PDF of $x_k$ and $z_k$ conditioned on $Z_{k-1}$. The PC-CRLB differs from the PCRLB in that it characterizes performance conditioned on the actual data that has been received.

The following are results from the real-time operation of a cognitive pulse-Doppler radar tracking system [Smi15], [Smi16]. The system tracks the range, velocity, and signal-to-interference-plus-noise (SINR) of a target. The cognitive algorithm adjusts the number of pulses (Np) and the PRF to meet performance goals for the range and velocity tracks, to keep the target out of the clutter, and to prevent Doppler ambiguity.

In the first experiment the PRF and Np were fixed at 6 kHz and 128 pulses, respectively. For this "baseline" experiment there was no adaptation of radar parameters. Fig. 8 shows the output results after running the experiment. The range, velocity, and SINR track plots all show the mean and two-sigma error of the predicted values along with the actual measurements. The normalized Doppler frequency (normalized to the PRF) plot shows the upper and lower bounds set for the PRF along with the maximum predicted and mean target Doppler values. The velocity and range standard deviation plots show the predicted and actual tracker root mean square error (RMSE) as a function of elapsed time, where the predicted RMSE is obtained from the PC-CRLB. The dotted line indicates the performance goal. Below that are the time elapsed plots for the PRF and number of pulses, which, in this first case, were fixed.

The high PRF value kept the measurement and tracking processes unambiguous. This was indicated by the magenta curve on the normalized Doppler frequency graph remaining below the upper bound. The target was generally kept out of the clutter, which is indicated by the green curve remaining above the lower bound. However, because of the high PRF, the Doppler filter bins were also wide and there was not enough Doppler resolution to control the velocity RMSE. Consequently, the predicted and actual velocity error significantly exceeded the error goal on two occasions. Additionally, fluctuations in the SINR caused the range error to similarly miss its goal. The degree of error in range was observed to be a fairly direct function of SINR. This would be deemed a non-optimum solution even for a traditional fixed parameter approach to radar design.

**Figure 8** – Experimental results for a fixed PRF (6kHz) and number of pulses (128).

The results of the second experiment, cognitively adapting the PRF and number of pulses, are shown in Fig. 9. It can be seen that the maximum predicted Doppler, as shown by the magenta trace, was always kept within the unambiguous limit, and the mean Doppler, as shown by the green trace, was always kept above the clutter limit. The changing PRF value can be seen directly in the bottom center figure. The driving of the PRF to the lowest possible value at any instant in time had the desired effect of controlling the velocity error RMSE so that it barely exceeded the prescribed performance level. This time, on the two occasions where the SINR started to fall, the number of integrated pulses was increased so that the effect on the range error as well as the velocity error was controlled and the achieved performance in the red curve was very close to that predicted in the green curve. The plots showing the velocity and range errors now show a performance in velocity that is within the specified performance limit with range just exceeding it on a couple of occasions.

**Figure 9** – Experimental results for cognitively adapted PRF and number of pulses

These results show the power of a cognitive approach to radar sensing for maintaining a desired level of tracking performance. Indeed, instead of designing a radar by specifying the system parameters and then developing sophisticated tracking algorithms to cope with Doppler ambiguities and target fading, we are now able to design a cognitive tracking radar by specifying the performance and allowing the system to set the transmit waveform parameters adaptively to achieve the desired performance.

The approach can be applied to sensor/processor systems involved not only in tracking, but also detection, localization, classification, and imaging.

The appropriate processor cost function would need to be specified, and the predicted conditional Bayes risk evaluated. In most problems, the predicted conditional Bayes risk is intractable to evaluate and it would need to be replaced by a suitable surrogate information measure. For example, joint tracking and classification was considered in [Kre05a] and the information measure used was expected Renyi divergence.

## 4. Cognitive passive radars

A Passive Radar (PR) system is a bistatic radar that makes use of emissions from a non co-operative transmitter of opportunity, such as broadcast, communications, or radio-navigation transmitters rather than a dedicated, co-operative radar transmitter. Such systems have a number of potential advantages over conventional active systems. The receiver is passive and so potentially undetectable. Many illumination sources can be used, and many of them are high power and favorably sited. PR receiver systems can often be rather simple and low cost, and there is no need for any license for the transmitter. Moreover, in recent years, multistatic PR systems have become very attractive for harbor protection and coastal surveillance, offering a number of advantages in terms of eco-compatibility and sustainability. In fact, they can be installed even in protected and populated areas, reasonably without providing additional electromagnetic (EM) pollution. In this sense, the use of PRs for low/medium range applications can be viewed as a strategy for a smart use of the spectrum resources [Gri15].

Passive radars are quite often grouped in networks to extend coverage and improve detection, tracking, and identification of targets entering the region under surveillance. This can be done easily even with a single receive node exploiting the different sources of illumination available in the surveillance area and the spatial diversity provided by different channels of observation. Clearly, passive radars cannot change in a cognitive way their transmitted waveform, because they do not transmit but fully rely on the sources of opportunity available in the surveillance area. However, if the receive node is able to handle multiple signals (FM, UMTS, DVB-T etc.) [DiL16], [O'Ha12], it can decide in a cognitive manner which channel or set of channels to use for detecting, tracking, and classifying the targets, based on the acquired information on targets themselves and knowledge of source characteristics and transmitter-target-receiver geometry, in the same way as an active radar chooses the transmit waveform on-the-fly, as shown in the previous section.

The example we describe here was first introduced in [Gre11], [Sti13], [Gre16]. It concerns the selection of a source of opportunity for tracking a target in a passive network where there is only a single receiver and multiple transmitters. The cognitive sensor selection algorithm is investigated with reference to a multistatic PR system sited in Leghorn harbor that exploits the signal emitted by two transmitters of opportunity: a UMTS Base Station and an FM commercial radio station. The FM station has some significant advantages because broadcast transmissions at VHF/UHF usually have substantial transmit powers, so they can have excellent coverage. Different advantages are achieved utilizing the UMTS sources of illumination. In fact, the FM system could detect targets at relatively far ranges and use its fine Doppler resolution to build a crude track. Once the target appears within the coverage area of the UMTS system, it can be exploited for its superior range resolution properties. In particular, the analyzed scenario, as shown in Fig. 10, is composed of one receiver and two transmitters. The receiver is placed in the Leghorn harbor. Its antenna has gain $G$=10dB and Half Power Beam Width (HPBW)=3°. The first transmitter, a UMTS Base Station, is located 453 m away from the receiver in the South-East direction. The second transmitter, an FM commercial radio station, is located on "Monte Serra", 36 km away from the receiver, in the North-East direction.

**Figure 10** - Multistatic PR System.

We consider the case in which the target trajectory is purely deterministic and the radar measurements at discrete-time $k$ are available only if the target has been detected. In particular, the target trajectory is described by the state vector $\mathbf{x}_k=[x_k, \dot{x}_k, y_k, \dot{y}_k]^T$, where $(x_k, y_k)$ is the position of the target at time $k$, while $(\dot{x}_k, \dot{y}_k)$ are the velocity components along the main axes of the coordinate system. Assuming that the target is moving with constant velocity and that the evolution of the state vector is deterministic, it is possible to write $\mathbf{x}_{k+1}=\mathbf{F}\mathbf{x}_k$, where $\mathbf{F}$ is the target model matrix. The objective of target tracking is to estimate recursively the target state from a set of measurements $\mathbf{z}_k=[r_k, v_k]^T$, that is the range from receiver to target and the bistatic velocity. The measurement equation can be put in the following vector form $\mathbf{z}_k = \mathbf{h}(\mathbf{x}_k) + \mathbf{w}_k$, where $\mathbf{h}$ is a known (non-linear) function of the state vector and $\mathbf{w}_k$ is a measurement noise sequence.

As stated before, the radar measurement is available at time $k$ only if the target has been detected and hence with a probability $P_D$, i.e. the probability of detection, which is itself dependent on the local signal-to-noise power ratio at time $k$, $SNR_k$. The bistatic measurements are affected by additive noise $\mathbf{w}_k$ that is assumed to be Gaussian distributed with zero mean and covariance matrix $\mathbf{R}_k$. In this case, $\mathbf{R}_k$ is equal to the Cramér-Rao Lower Bound (CRLB) of the target velocity and range. The tracking system in the receiver node is assumed to know the CRLBs over the entire area under surveillance for both UMTS and FM signals, which depend on the bistatic geometry and on the exploited transmitted waveform [Gre11], [Gre16], [Sti13].

For the model assumed here, the PC-CRLB and PCRLB are the same and do not depend on the actual sequence of observed measurements, but only on the measurement covariance matrix $\mathbf{R}_k$. Furthermore, since the target state is deterministic, the PC-CRLB and PCRLB reduce to the bistatic CRLB. Let us denote by $\mathbf{J}_k$ the filtering information matrix of the state vector at time $k$. Its inverse is the bistatic CRLB that bounds the error variance of the target state estimate at time $k$, that is $\mathbf{J}_k^{-1} \leq E\{(\hat{\mathbf{x}}_{k|k} - \mathbf{x}_k)(\hat{\mathbf{x}}_{k|k} - \mathbf{x}_k)^T\}$ where $\hat{\mathbf{x}}_{k|k}$ is an unbiased estimator of the state vector based on all the available measurements up to time $k$. Using the results shown in [Ris04b], [Tic98], it is possible to demonstrate that $\mathbf{J}_k$ can be computed recursively using the following equation:

$$\mathbf{J}_{k+1} = \left[\mathbf{F}^{-1}\right]^T \mathbf{J}_k \mathbf{F}^{-1} + P_D \mathbf{H}_{k+1}^T \mathbf{R}_{k+1}^{-1} \mathbf{H}_{k+1}, \quad (3)$$

where $\mathbf{H}_{k+1}$ is the Jacobian of $\mathbf{h}(\mathbf{x}_{k+1})$ evaluated at the state $\mathbf{x}_{k+1}$. The recursion in (3) starts with the initial Fisher Information Matrix (FIM) $\mathbf{J}_0$ that, assuming the initial distribution of $\mathbf{x}_0$ is Gaussian, is equal to the inverse of the covariance matrix of $\mathbf{x}_0$.

The recursive equation in (3) is the sum of two terms: the first one $[\mathbf{F}^{-1}]^T \mathbf{J}_k^{-1} \mathbf{F}^{-1}$ is the a priori information given by the previous target state, while $P_D \mathbf{H}^T_{k+1} \mathbf{R}^{-1}_{k+1} \mathbf{H}_{k+1}$ is the information gained by the radar measurements. The term $P_D$ can be intuitively justified considering that there is a measurement information contribution only if the target has been detected.

As shown, the CRLB depends on the target trajectory, sensor accuracy, transmitted waveform, probability of detection, and bistatic geometry. This dependence is given by the measurement information term, which is itself dependent on the probability of detection $P_D$ and on the measurement covariance matrix $\mathbf{R}_k$. In particular, the target dependent quantity $P_D \cdot \det(\mathbf{R}_k^{-1})$ can be considered as a measure of the measurement information, that is, the higher $P_D \cdot \det(\mathbf{R}_k^{-1})$, the higher the information gained by the radar measurement. Therefore, this quantity can be used to select the channel with the best performance for each point of the analyzed area.

Knowing the transmitter positions and the estimated state vector of the target at each time $k$, the receiver is able to calculate, for each point of the target trajectory, the quantity $P_D \cdot \det(\mathbf{R}_k^{-1})$ and therefore it is able to evaluate the channel with the best performance.

The target trajectory being deterministic, it is easy to verify that the CRLBs of the receiver that dynamically selects the best channel are given by:

$$\mathbf{J}_{k+1} = \left[\mathbf{F}^{-1}\right]^T \mathbf{J}_k \mathbf{F}^{-1} + \mathbf{J}_{k+1}^{CH} \quad (4)$$

where

$$\mathbf{J}_{k+1}^{CH} = \begin{cases} P_D^{(1)} \left[\mathbf{H}_{k+1}^{(1)}\right]^T \left[\mathbf{R}_{k+1}^{(1)}\right]^{-1} \mathbf{H}_{k+1}^{(1)}, & \text{if } P_D^{(1)} \det\left(\left[\mathbf{R}_{k+1}^{(1)}\right]^{-1}\right) > P_D^{(2)} \det\left(\left[\mathbf{R}_{k+1}^{(2)}\right]^{-1}\right) \\ P_D^{(2)} \left[\mathbf{H}_{k+1}^{(2)}\right]^T \left[\mathbf{R}_{k+1}^{(2)}\right]^{-1} \mathbf{H}_{k+1}^{(2)}, & \text{if } P_D^{(1)} \det\left(\left[\mathbf{R}_{k+1}^{(1)}\right]^{-1}\right) < P_D^{(2)} \det\left(\left[\mathbf{R}_{k+1}^{(2)}\right]^{-1}\right) \end{cases}$$

(5)

that is, at each step $k+1$ the measurement information term $\mathbf{J}^{CH}_{k+1}$ is given by selecting the channel with the higher measurement information. The upper index "1" refers to the UMTS channel while the index "2" to the FM channel.

The CRLBs of the receiver that selects the best transmitter are shown in Figs. 11-12. From the results it is evident that there is a substantial gain with respect to each bistatic channel and the obtained performance is equal to or better than the performance of the channel with the lowest CRLB.

As evident in the figures, initially the performance is the same as that of the FM channel. This is due to the fact that in the far range the FM channel has the best performance thanks to the higher $SNR$. When the target approaches the harbor, the UMTS channel has better resolution and, exploiting this channel, the proposed receiver is able to improve the performance in estimating the target trajectory [Sti13].

**Figure 11.** Root of the PCRLB of target state. UMTS channel, FM channel and dynamic selection channel; x coordinate.

**Figure 12.** Root of the PCRLB of target state. UMTS channel, FM channel and dynamic selection channel; y coordinate.

Active and passive technologies can be used jointly, as proposed in [Stin17] where a new idea of "Symbiotic" Radar (SR) is introduced. The proposed symbiotic radar is integrated with an IEEE 802.22 WRAN (Wireless Regional Area Network). The radar receiver is based on passive radar technology and exploits the IEEE 802.22 devices as transmitters of opportunity. But, being integrated with the WRAN, it can control the Medium Access Control (MAC) layer of the WRAN, selecting, in a collaborative way with the base station (BS), some of the Customer Premise Equipments (CPEs). These selected CPEs are scheduled to transmit in each frame to improve radar performance (active technology), based again upon the knowledge of the environment.

The IEEE 802.22 is a new standard based on Cognitive Radio techniques for WRAN that exploits, in a non-interfering and opportunistic basis, the unused channels in the VHF and UHF bands allocated to television. The architecture of the IEEE 802.22 network is composed of a base station that covers a cell with a radius up to 30 km, providing high-speed internet service for $N$ CPE devices or groups of devices up to 512. All the details on the IEEE 802.22 standard can be found in [STD11]. Some recent works (see for instance [Mis15], [Sti16c]) analyzed the possibility to exploit IEEE 802.22 devices as transmitters of opportunity for purely passive radar systems.

A symbiotic radar system is not purely passive, it does not simply select the best sources of opportunity among those available, based upon some optimality criterion. A symbiotic radar, as conceived in [Sti17] can ask a specific CPE to transmit (if silent) in order to improve the detection or tracking performance of the radar. The signal transmitted by the CPEs always follows the IEEE 802.22 standard and the overall system should guarantee a preset Quality of Service (QoS) for communications. Therefore, in the symbiotic radar the communication and the radar system work cooperatively.

Supposing that all the $N$ CPEs are transmitting to the BS, and that the passive radar would use $M$ ($<<N$) of those CPEs signals for its tracking function, the best set $S_{ideal}$ of $M$ CPEs could be chosen, for instance, in order to minimize the MSE of the target state prediction. In [Hay12] it has been proven that this is tantamount to minimizing the trace of the prediction of the target state covariance matrix $P_{k+1|k}$ at each time instant $k$, so the minimum value of the cost function is given by Trace$\{P_{k+1|k}(S_{ideal})\}$. Unfortunately, the radar can rarely use the set $S_{ideal}$ because only few CPEs transmit at the same time, so in the general purely passive case, the radar should select its set from among the few that are active at the instant $k$. In the symbiotic radar, conversely, the BS can choose $n$ of the $M$ sources from among those that are not transmitting, requiring them to transmit in order to improve the tracking performance. The cognitive tracking algorithm selects the set of transmitters for the subsequent frame by finding the minimum number $n$ that guarantees a performance level such that

$$\lambda \cdot Trace\{\mathbf{P}_{k+1|k}(S_n)\} \leq Trace\{\mathbf{P}_{k+1|k}(S_{ideal})\} \quad (6)$$

where $0 \leq \lambda < 1$. The threshold $\lambda$ is used to select how the desired performance compares with that of the ideal case. When $\lambda=0$, the tracking is purely passive and the SR does not select any CPE. In this case the SR will not improve the target tracking performance. On the other hand, when $\lambda$ tends to 1, the SR improves tracking performance as closely to the ideal case as possible.

Some results of this approach are shown in Figs. 13 and 14. The target is moving linearly on a plane from [-400 m, 400 m] to [400 m, -250 m], with a speed of 8.33 m/sec. The gain of the symbiotic approach compared to the purely passive one is evident in Fig. 13 where the RMSE of an Extended Kalman Filter (EKF) on the target position is plotted as a function of the time instant $k$. Comparing the black curve ($\lambda=0$) and the blue curve ($\lambda=0.95$), the symbiotic approach obtains performance nearly identical to the ideal set. In this example $N=256$, $M=8$ and $n$ goes from 0 to a maximum of 4. In Fig. 14, the number $n$ of CPEs selected specifically by the SR is plotted as a function of the time instant $k$, calculated on $10^4$ Monte Carlo trials. It is evident that, for the majority of the time, one or two selected CPEs are enough to guarantee performance close to the ideal one.

**Figure 13** - RMSE of target position.

**Figure 14** - Mean number of CPEs scheduled to transmit by the SR.

## 5. Open issues and future research directions

Although fully cognitive radars, inspired by the bat and dolphin bio-sonars, as conceived in [Hay06], are not yet a reality, some encouraging attempts to add cognition to classical radar systems have been accomplished, both theoretically and practically, as presented in this article and proven by the many conference and journal papers published over the last decade (a sample of them are in the list of references). However, many problems are still open and require additional effort to obtain solutions to make cognitive radars a reality.

In active radars, cognition requires waveforms and circuits to be reconfigurable and optimizable in real time. Initial progress has been made in the two separate fields [Bay12] but a fully optimized solution that includes all the important aspects of radar circuitry has not yet been presented [Bay14] even though some attempts to consider the radar as a holistic system (hardware-in-the-loop) have been presented, for instance, in [Jak12].

The dynamic reconfiguration of the spectrum portion to be used for transmitting, as described in previous sections, is not always easily implementable. The main reason is that quite often, due to the non-linear operational regime of the high-power radar RF circuitry (particularly for vacuum tube amplifiers), there is a non-negligible spectral spreading outside the assigned radar band (spectral regrowth). This makes coexistence of communications and radar systems in close bands with narrow guard bands difficult [Gri15]. Magnetron tubes, quite often used in legacy radar systems because they are inexpensive, have serious drawbacks in term of spectral purity. To reduce the out-of-band (OOB) emissions, bandpass filters are often used, though the cost of this improvement in spectral purity means a significant loss in the effective transmitted power.

Solid-state-based amplifiers are much easier to control in terms of OOB, but unfortunately, they cannot provide the high peak power of tubes and, anyway, they represent only a small minority of current operational systems.

Of course, the frequency use and emissions by radars and other transmitting devices are all regulated. Many countries, but not all, adopt the ITU emission standard [ITU12]. Fig. 15 shows a typical emission mask that might be applied to radar systems.

**Figure 15** – Graph of a generic ITU spectral mask, showing the required suppressions relative to power at fundamental (dB)

There is a band over which the radar is designed to transmit. It is fixed in frequency and goes down -40 dB from the peak. Outside, at lower power levels, OOB emissions are permitted with, generally, a roll-off of -20dB/decade (-40db/decade is under consideration). The radar transmissions should not exceed the limits imposed by the mask, but unfortunately unwanted emissions, due to nonlinearity in the transmitter and to the steep rise and fall times of the radar pulses, often occur [Gri15].

An intermediate step toward arbitrary waveform generation is selection of waveforms or waveform parameters from a pre-specified set. Many modern radars already have this capability and a first step toward making cognitive radars a reality could be implementing cognitive processing to choose among the set of allowable waveforms [Blu16].

In passive multisensory radar systems, the cost must be kept low, because this is one of the main reasons that justify their use, despite their poorer performance compared to active systems. Cognitive algorithms implemented on passive systems should then be easy to implement, and not be very demanding in terms of energy and memory usage. Fortunately, the rapid increase in the performance of DSPs, FPGAs and ASICs have made the signal processing more compact and low power [Ing10].

There are numerous exciting future research directions to be explored to make cognitive radars a reality. Some general areas include extension of the basic concepts to multi-user and multi-objective systems and to systems with large degrees of freedom available for adaptation (e.g. frequency, antennas, waveforms/codes, polarization, power resources, transmitter/receiver selection, etc.) and expansion of the role of learning and knowledge storage/recovery over longer time horizons.

What is really difficult to implement in a "machine" such as a radar is the capability of learning from the mistakes that occurred as a result of poor decisions in the past, and hence the ability to make a very informed decision in the future, as envisioned in [Hay06].

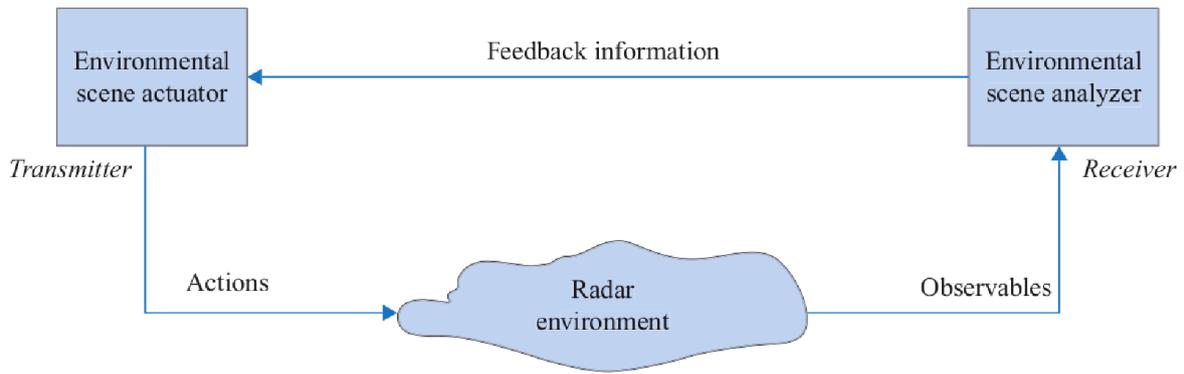

Figure 1 - Block diagram of cognitive radar seen as a dynamic closed-loop feedback system with the perception-action cycle [Hay12].

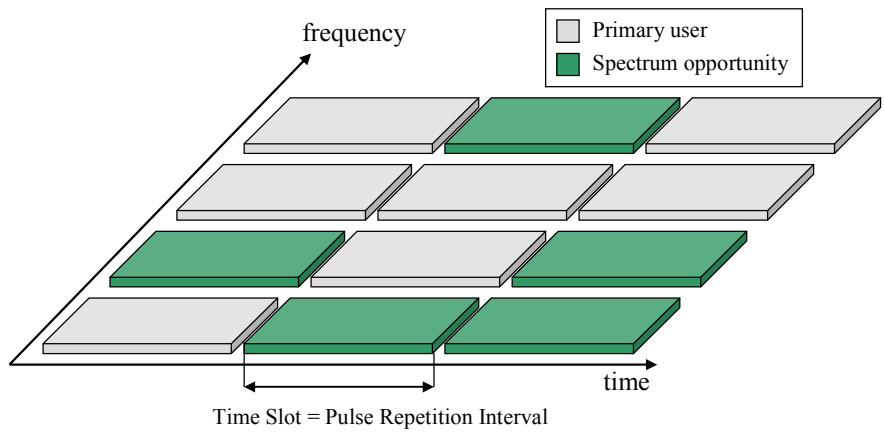

**Figure 2** – Spectrum opportunities [Gre16].

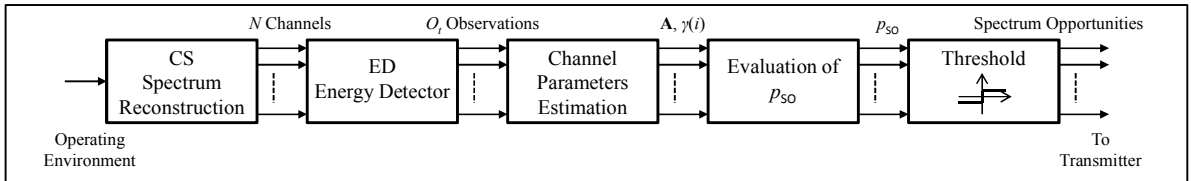

**Figure 3** – Blocks in a cognitive radar.

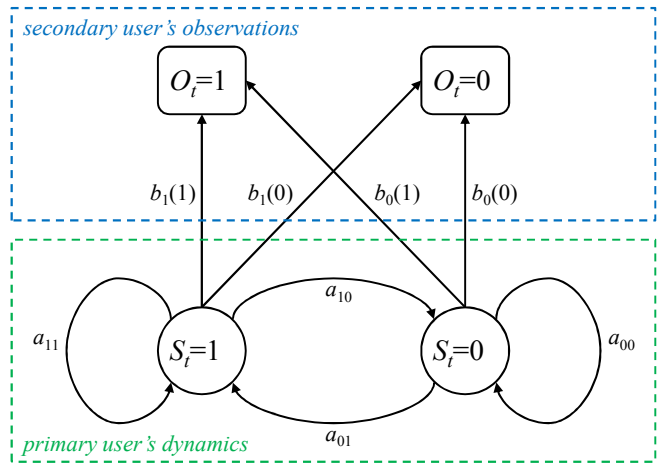

**Figure 4** - Hidden Markov Model representation for spectrum occupancy.

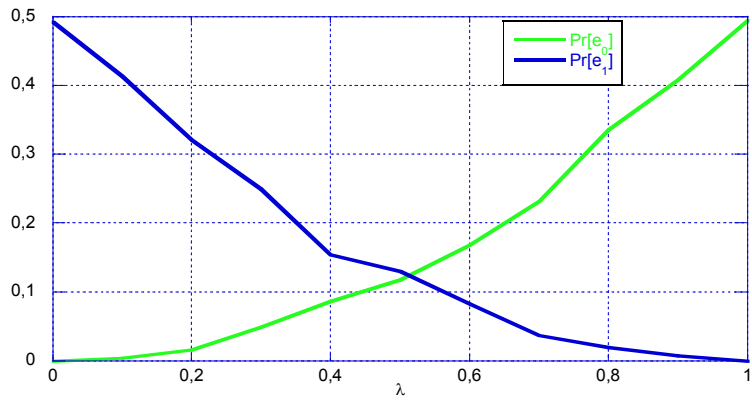

**Figure 5** – Probabilities of $e_0$ and $e_1$ as a function of $\lambda$ [Sti16a].

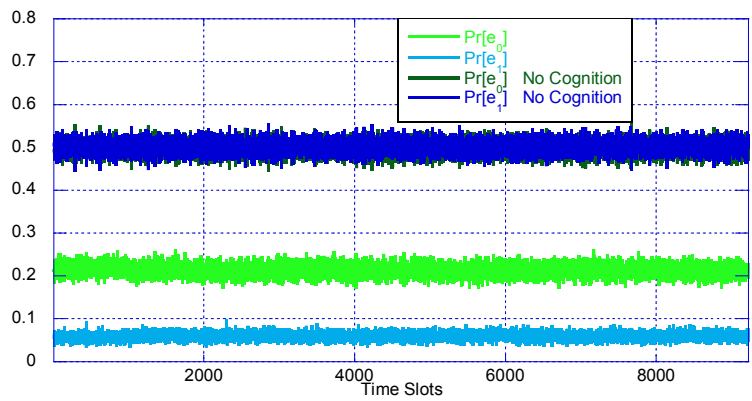

**Figure 6** - Probabilities of $e_0$ and $e_1$ as a function of time [Sti16a].

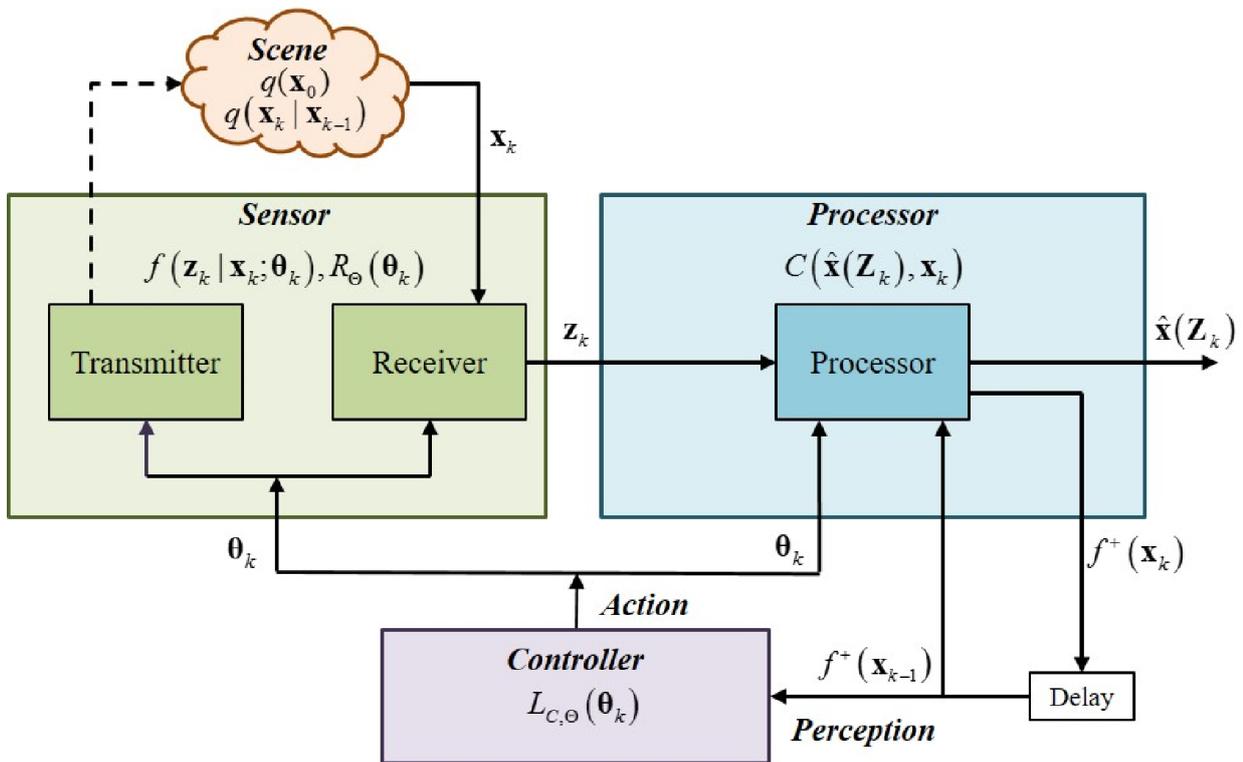

**Figure 7** – Cognitive Sensor/Processor System

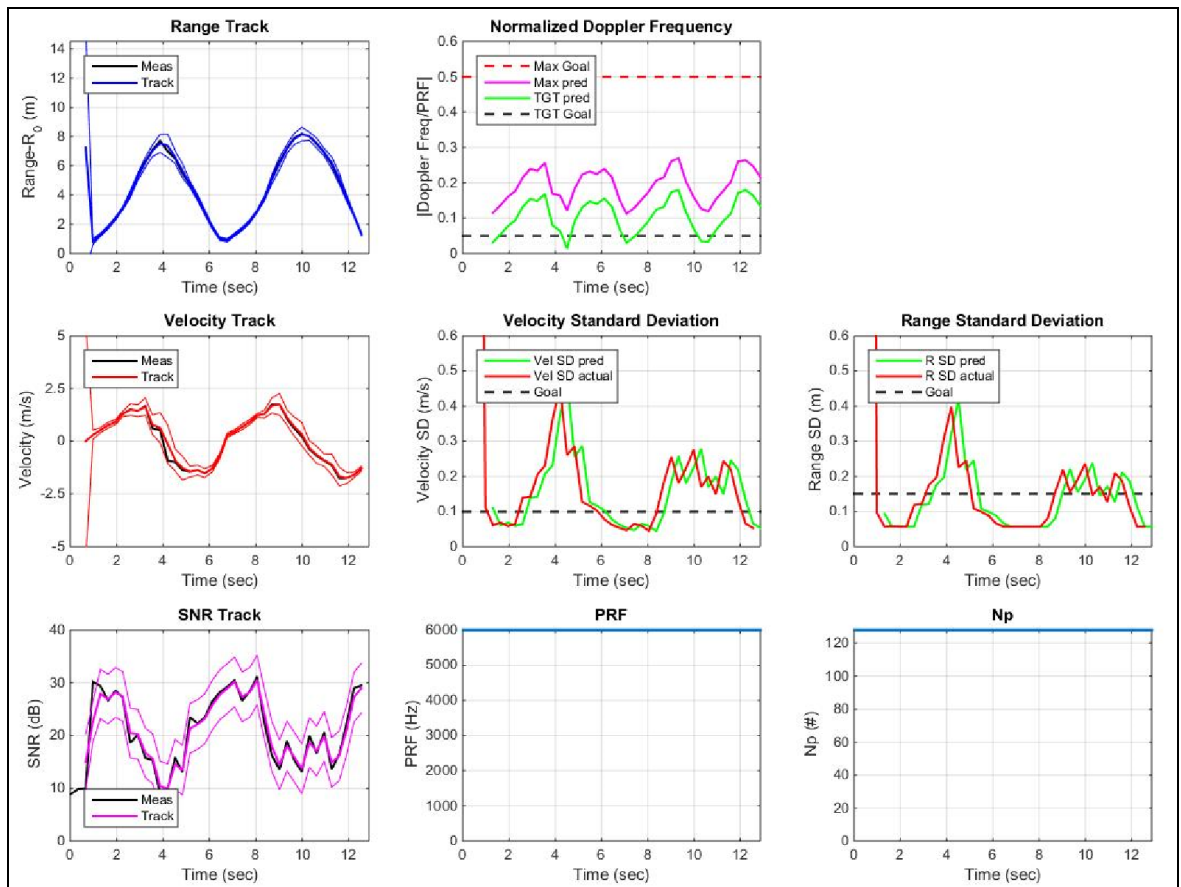

**Figure 8** – Experimental results for a fixed PRF (6kHz) and number of pulses (128).

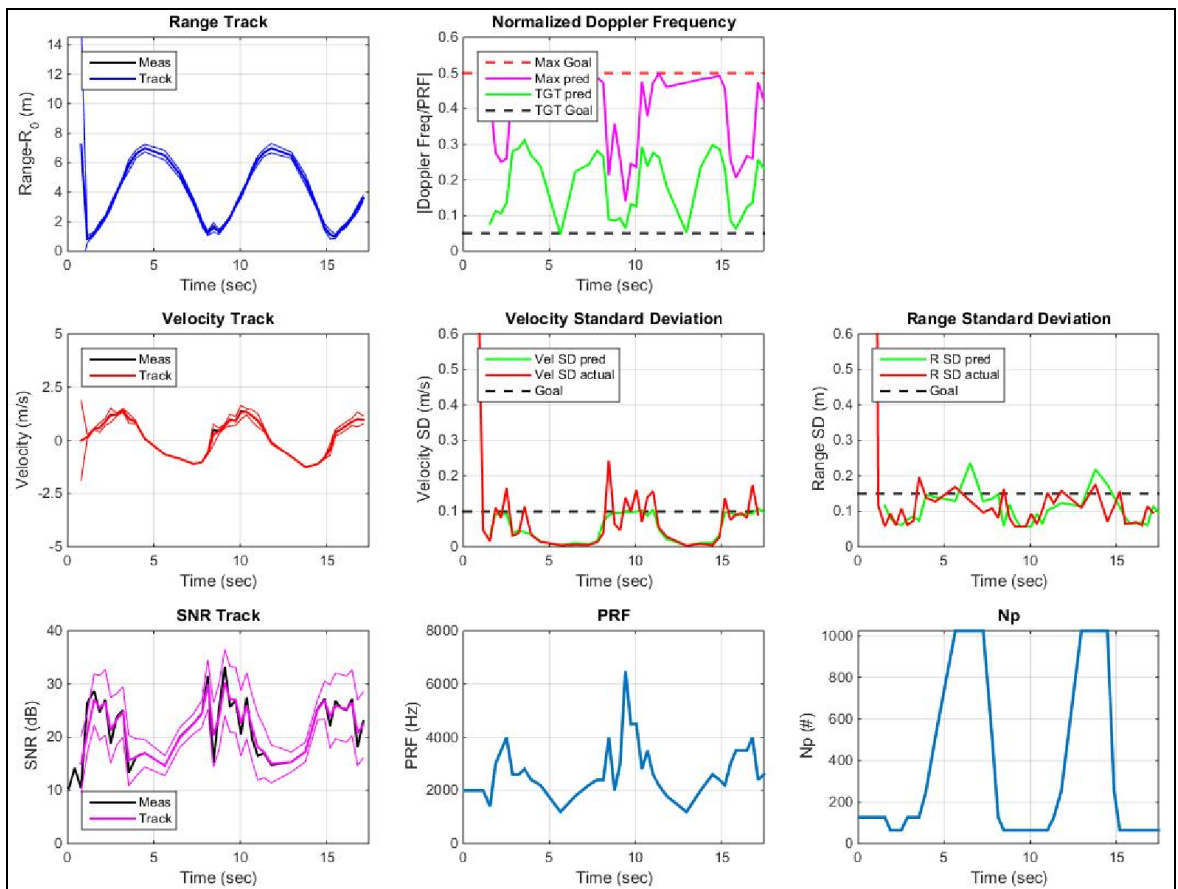

**Figure 9** – Experimental results for cognitively adapted PRF and number of pulses

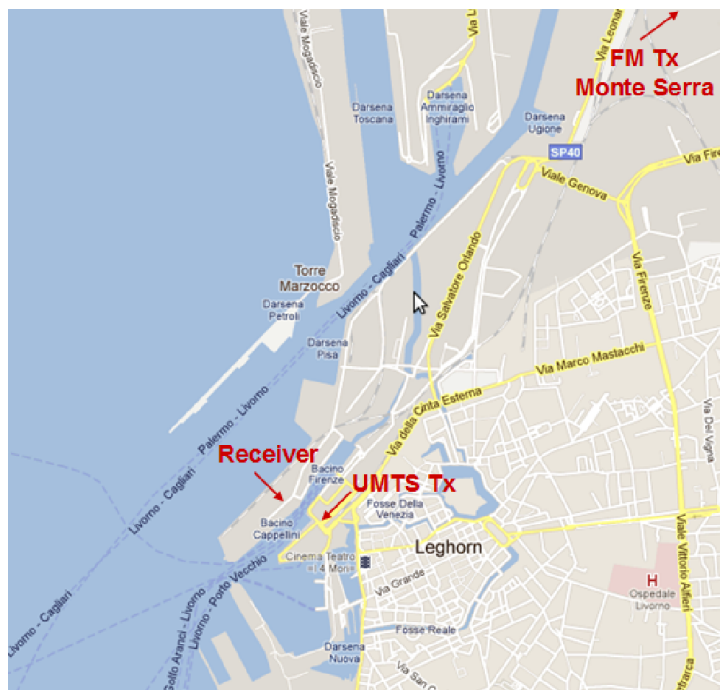

**Figure 10** - Multistatic PR System.

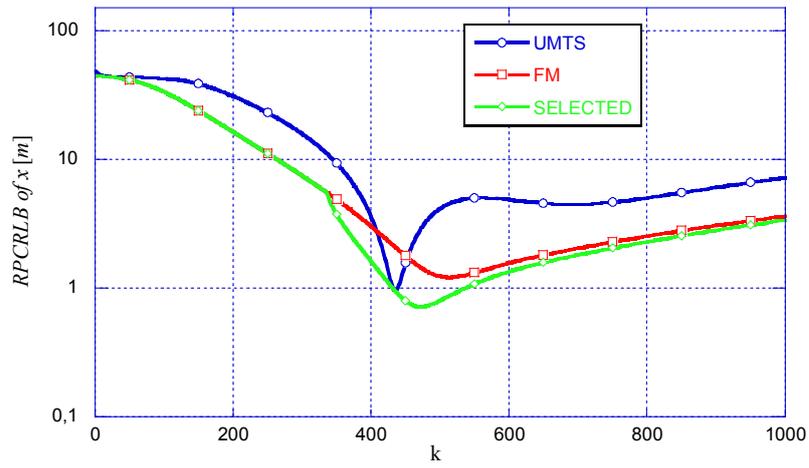

**Figure 11.** Root of the PCRLB of target state. UMTS channel, FM channel and dynamic selection channel; *x* coordinate.

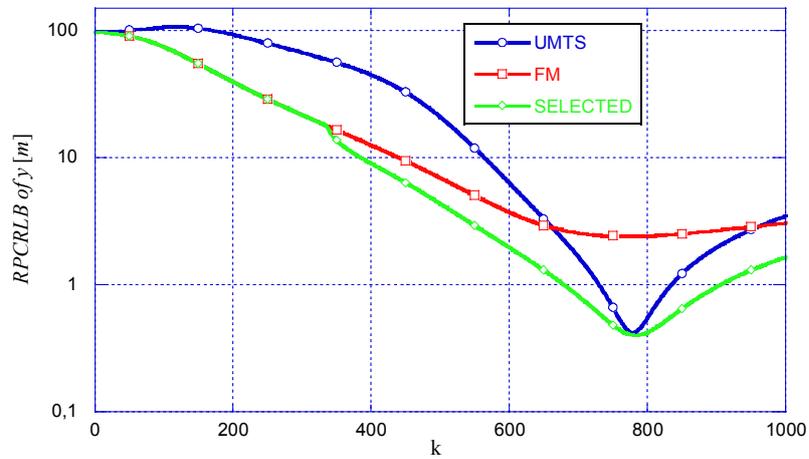

**Figure 12.** Root of the PCRLB of target state. UMTS channel, FM channel and dynamic selection channel; y coordinate.

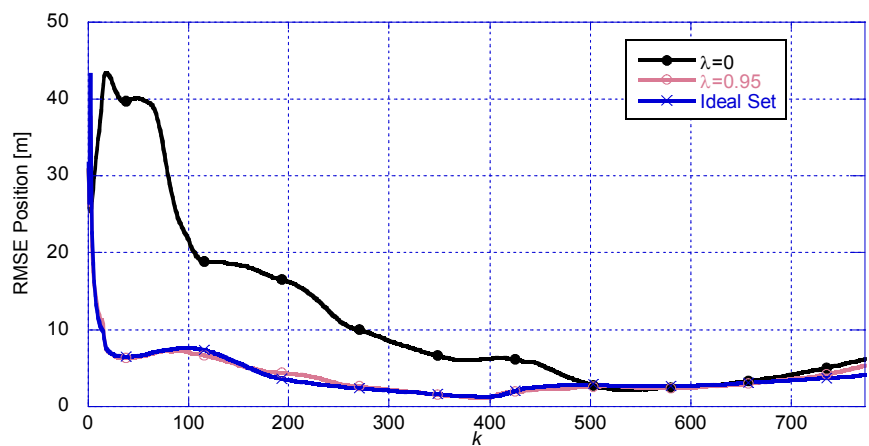

**Figure 13** - RMSE of target position.

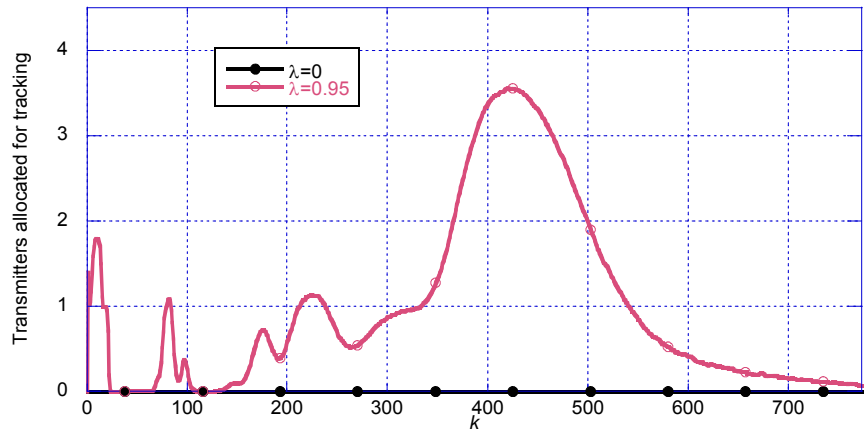

**Figure 14** - Mean number of CPEs scheduled to transmit by the SR.

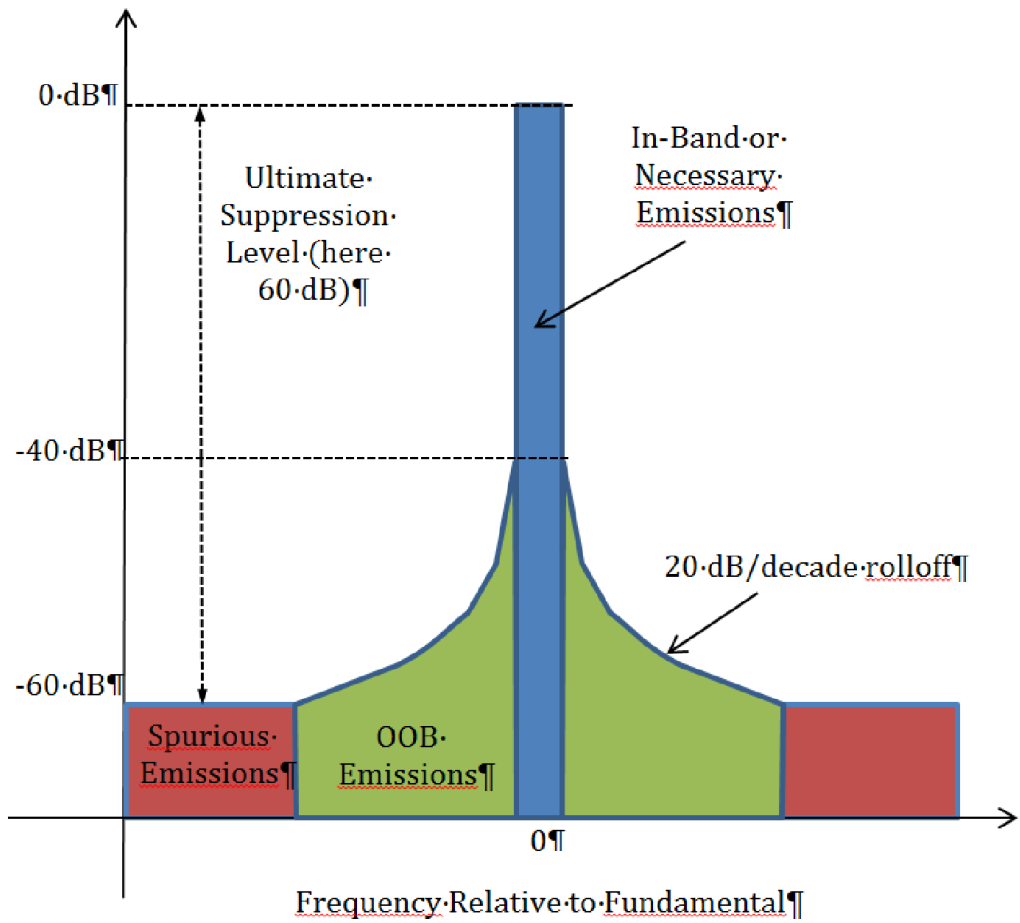

**Figure 15** – Graph of a generic ITU spectral mask, showing the required suppressions relative to power at fundamental (dB)